\documentclass[article]{JHEP3}

\usepackage{amsmath,amssymb}
\usepackage[dvips]{graphics}
\usepackage{epsfig}
\usepackage{psfrag}

\newcommand{\be}{\begin{equation}}
\newcommand{\ee}{\end{equation}}
\newcommand{\bea}{\begin{eqnarray}}
\newcommand{\eea}{\end{eqnarray}}
\newcommand{\bean}{\begin{eqnarray*}}
\newcommand{\eean}{\end{eqnarray*}}

\relax

\def\MM{\boldsymbol{\mu}}

\def\LL{\boldsymbol{\lambda}}

\preprint{\small \texttt{hep-th/0403132}}

\title{Asymptotic Quasinormal Frequencies for Black Holes in Non--Asymptotically Flat Spacetimes}

\author{Vitor Cardoso$^{\dag}$, Jos\'e Nat\'ario$^{\ddag}$ and Ricardo
Schiappa$^{\ddag,\dag\dag}$
\\
$^{\dag}$Centro de F\'{\i}sica Computacional, Universidade de Coimbra,\\ 
P--3004--516 Coimbra, Portugal\\
\\
$^{\ddag}$CAMGSD, Departamento de Matem\'atica, Instituto Superior T\'ecnico,\\ 
Av. Rovisco Pais 1, 1049--001 Lisboa, Portugal\\
\\
$^{\dag\dag}$Faculdade de Engenharia, Universidade Cat\'olica 
Portuguesa,\\
Estrada de Tala\'\i de, 2635--631 Rio de Mouro, Lisboa, Portugal\\
\\
\email{vcardoso@fisica.ist.utl.pt}, \quad
\email{jnatar@math.ist.utl.pt}, \quad 
\email{schiappa@math.ist.utl.pt}
}

\abstract{
The exact computation of asymptotic quasinormal frequencies is a technical problem which involves the analytic continuation of a Schr\"odinger--like equation to the complex plane and then performing a method of monodromy matching at the several poles in the plane. While this method was successfully used in asymptotically flat spacetime, as applied to both the Schwarzschild and Reissner--Nordstr\o m solutions, its extension to non--asymptotically flat spacetimes has not been achieved yet. In this work it is shown how to extend the method to this case, with the explicit analysis of Schwarzschild de Sitter and large Schwarzschild Anti--de Sitter black holes, both in four dimensions. We obtain, for the first time, analytic expressions for the asymptotic quasinormal frequencies of these black hole spacetimes, and our results match previous numerical calculations with great accuracy. We also list some results concerning the general classification of asymptotic quasinormal frequencies in $d$--dimensional spacetimes.
}


\begin{document}


\section{Introduction}


A long time has passed since research first focused on analyzing the linear stability of four dimensional black hole solutions in general relativity \cite{regge-wheeler, zerilli-1}. However, it was not until very recent times that this stability problem was addressed within a $d$--dimensional setting \cite{kodama-ishibashi-1, kodama-ishibashi-2, kodama-ishibashi-3}. These papers tried to be as exhaustive as possible, studying in detail the perturbation theory of spherically symmetric black holes in $d$--dimensions and allowing for the possibilities of both charge and a background cosmological constant. Having thus acquired a list of stable black hole solutions, the next question to address within this problem are quasinormal modes---the damped oscillations which describe the return to the initial configuration, after the onset of a linear perturbation (see \cite{nollert, kokkotas-schmidt} for reviews). 

Besides their natural role in the perturbation theory of general relativity, quasinormal modes have recently been focus of much attention following suggestions that they could have a role to play in the quest for a theory of quantum gravity \cite{hod, dreyer}. The idea is to look at those special modes which are infinitely damped, and thus do not radiate. It was suggested in \cite{hod} that an application of Bohr's correspondence principle to these asymptotic quasinormal frequencies could yield new information about quantum gravity, in particular on the quantization of area at a black hole event horizon. It was further suggested in \cite{dreyer} that asymptotic quasinormal frequencies could help fix certain parameters in loop quantum gravity. Both these suggestions lie deeply on the fact that the real part of the asymptotic quasinormal frequencies is given by the logarithm of an integer number, a fact that was analytically shown to be true, for Schwarzschild black holes in $d$--dimensional spacetime, in \cite{motl, motl-neitzke}. A question of particular relevance that immediately follows is whether the suggestions in \cite{hod, dreyer} are universal or are only applicable to the Schwarzschild solution. Given the mentioned analysis of \cite{kodama-ishibashi-1, kodama-ishibashi-3}, one has at hand all the required information to address this problem and compute asymptotic quasinormal frequencies of $d$--dimensional black holes. A preliminary clue is already present in \cite{motl-neitzke}, where the analysis of the four dimensional Reissner--Nordstr\o m solution yielded a negative answer: the asymptotic quasinormal frequencies obeyed a complicated relation which did not seem to have the required form. While extending this result to both the $d$--dimensional and the extremal  Reissner--Nordstr\o m solutions did not pose great obstacles \cite{natario-schiappa}, an extension of the analytical techniques in \cite{motl-neitzke} to non--asymptotically flat spacetimes proves to be a greater challenge. It is the goal of this paper to carry out an extension of the techniques in \cite{motl-neitzke} to non--asymptotically flat spacetimes, with the explicit analysis of Schwarzschild de Sitter and large Schwarzschild Anti--de Sitter black holes, both in four dimensions. The detailed study of these solutions in $d$--dimensions will appear elsewhere \cite{natario-schiappa}, including charged solutions in asymptotically de Sitter and asymptotically Anti--de Sitter spacetimes, as well as an analysis of the implications of our results on what concerns the proposals of \cite{hod, dreyer}, dealing with the application of quasinormal modes to quantum gravity.

It is important to stress that even if the ideas in \cite{hod, dreyer} turn out not to be universal, it is still the case that quasinormal frequencies will most likely have a role to play in the quest for a theory of quantum gravity. Indeed, quasinormal frequencies can also be regarded as the poles in the black hole greybody factors which play a pivotal role in the study of Hawking radiation. Furthermore, the monodromy technique introduced in \cite{motl-neitzke} to analytically compute asymptotic quasinormal frequencies was later extended, in \cite{neitzke}, so that it can also be used in the computation of asymptotic greybody factors. It was first suggested in \cite{neitzke} that the results obtained for these asymptotic greybody factors could be of help in identifying the dual conformal field theory which microscopically describes the black hole, and these ideas have been taken one step forward with the recent work of \cite{krasnov-solodukhin}. It remains to be seen how much asymptotic quasinormal modes and greybody factors can help in understanding quantum gravity.

Let us conclude this introduction with some generics concerning quasinormal frequencies (we refer the reader to the upcoming \cite{natario-schiappa} for a full list of conventions and details). Later, in section 2, we shall compute asymptotic quasinormal frequencies for a Schwarzschild de Sitter black hole in four dimensional spacetime. Our results will also be shown to match earlier numerical computations with great accuracy. In section 3, we shall study large Schwarzschild Anti--de Sitter black holes in four dimensions, and analytically compute their asymptotic quasinormal frequencies. Again, our results match earlier numerical computations to great accuracy. We end with some comments concerning the general classification of asymptotic quasinormal frequencies in $d$--dimensional spacetimes \cite{natario-schiappa}.

For a four dimensional Schwarzschild black hole, one has the asymptotic quasinormal frequencies

$$
\lim_{n \to + \infty} \omega_{n} \sim [ {\mathrm{offset}} ] + i n [ {\mathrm{gap}} ] + {\mathcal{O}} \left( \frac{1}{\sqrt{n}} \right),
$$

\noindent
where the real part of the offset is the frequency of the emitted radiation, and the gap are the quantized increments in the inverse relaxation time. Here, the gap is given by the surface gravity. One can try to extend this analysis to more general situations and also include spacetimes with two horizons, but then generic results become much harder to obtain \cite{mmv-1, padmanabhan, mmv-2, choudhury-padmanabhan}. We shall take the time dependence for the perturbation to be $e^{i\omega t}$, so that ${\mathbb{I}}{\mathrm{m}} (\omega) > 0$ for stable solutions. There is also a reflection symmetry $\omega \leftrightarrow - \bar{\omega}$ which changes the sign of ${\mathbb{R}}{\mathrm{e}} (\omega)$. In this case, our quasinormal mode conventions are the following (see \cite{natario-schiappa} for a full list of conventions in $d$--dimensions). The perturbation master equations of \cite{kodama-ishibashi-1, kodama-ishibashi-3} can be cast in a Schr\"odinger--like form as 

\begin{equation} \label{schrodinger}
- \frac{ d^{2} \Phi_{\omega}}{dx^{2}} (x) + V (x) \Phi_{\omega} (x) 
= \omega^{2} \Phi_{\omega} (x),
\end{equation}

\noindent
where the potential will vary according to the specific case at hand. The boundary conditions are the usual: incoming waves at the black hole horizon and outgoing waves at infinity (or at the cosmological horizon, for the asymptotically de Sitter case)\footnote{For the asymptotically Anti--de Sitter situation things will be different.}. These can be written as

\begin{eqnarray*}
\Phi_{\omega} (x) &\sim& e^{i\omega x}\;\, {\mathrm{as}}\;\, x 
\to - \infty, \\
\Phi_{\omega} (x) &\sim& e^{-i\omega x}\;\, {\mathrm{as}}\;\, x 
\to + \infty,
\end{eqnarray*}

\noindent
where $x$ is the tortoise coordinate. Indeed, if the metric is chosen as $g = - f(r)\ dt \otimes dt + {f(r)}^{-1}\ dr \otimes dr + r^{2} d\Omega_{2}^{2}$, with parameters $M = \MM$ for the black hole mass and $\Lambda = 3 \LL$ for the background cosmological constant, then at any (event or cosmological) horizon, $f(R_{H})=0$. One can expand near the horizon $f(r) \simeq (r-R_{H}) f'(R_{H}) + \cdots$, and it follows for the tortoise

$$
x \equiv \int \frac{dr}{f(r)} \simeq \int \frac{dr}{(r-R_{H}) f'(R_{H})} = \frac{1}{f'(R_{H})} \log (r-R_{H}) \equiv \frac{1}{2k_{H}} \log (r-R_{H}) \equiv \frac{1}{4 \pi T_{H}} \log (r-R_{H}),
$$

\noindent
locally near the chosen horizon. Here $k_{H}$ is the surface gravity and $T_{H}$ is the Hawking temperature.


\section{Asymptotically de Sitter Spacetimes}


\cite{kodama-ishibashi-3} discusses the stability of black holes in asymptotically de Sitter (dS) spacetimes to tensor, vector and scalar--type perturbations of the metric and the electromagnetic field. For black holes without charge, which is the case we shall focus on, tensor and vector--type perturbations are stable in any dimension. Scalar--type perturbations are stable up to dimension six but there is no proof of stability in dimension $d \ge 7$. As we shall work in four dimensions, we are guaranteed a stable solution. Quantization in dS space was first addressed in \cite{gibbons-hawking}. These authors found that the cosmological event horizon is stable, but also that there is an isotropic background of thermal radiation. Analysis of the wave equation in dS space also led to the natural boundary conditions on quasinormal modes: incoming waves at the black hole horizon and outgoing waves at the cosmological horizon. The Schwarzschild dS solution in dimension $d=4$ has parameters $\MM$ and $\LL > 0$, with metric

$$
f(r) = 1 - \frac{2\MM}{r} - \LL r^{2}.
$$

\noindent
The potentials to be used in the master equation (\ref{schrodinger}), describing the evolution of scalar, electromagnetic and gravitational fields, can be followed through the Klein--Gordon, Maxwell and Einstein equations, respectively. They will necessarily depend on the specific field under consideration and are as follows. For scalar perturbations \cite{horowitz-hubeny}

\begin{equation} \label{vscalar}
V_{\mathsf{s}} (r) = f(r) \left( \frac{\ell \left( \ell+1 \right)}{r^2} + \frac{2\MM}{r^3} - 2 \LL \right),
\end{equation}

\noindent
while for electromagnetic perturbations \cite{cardoso-lemos-2}

\begin{equation} \label{vem}
V_{\mathsf{em}} (r) = f(r) \left( \frac{\ell \left( \ell+1 \right)}{r^2} \right).
\end{equation}  

\noindent
The gravitational perturbations decompose into two sets, the odd and the even parity one \cite{cardoso-lemos-2}. For the odd parity perturbations one has (these are the vector--type gravitational perturbations)

\begin{equation} \label{vodd}
V_{\mathsf{odd}} (r) = f(r) \left( \frac{\ell \left( \ell+1 \right)}{r^2} - \frac{6\MM}{r^3} \right),
\end{equation}

\noindent
while for the even parity perturbations (these are the scalar--type gravitational perturbations),

\begin{equation} \label{veven}
V_{\mathsf{even}} (r) = \frac{2f(r)}{r^3}\ \frac{9 \MM^3 + 3 a^2 \MM r^2 + a^2 \left( 1+a \right) r^3 + 3 \MM^2 \left( 3 a r - 3 \LL r^3 \right)}                                        
{\left( 3 \MM + a r \right)^2}
\end{equation}

\noindent
where $a = \frac{1}{2} \left( \ell (\ell+1) - 2 \right)$. In all cases, we have denoted by $\ell$ the angular momentum quantum number, which yields the multipolarity of the field. These are the potentials we shall use in the following.

To simplify the calculation, we choose the radius of the black hole to be our length unit. The radius of the cosmological horizon will then be an adimensional quantity $R>1$. In this case, it is easily seen that the warp factor must be of the form

$$
f(r)=1 - \frac{2\MM}{r} - \LL r^2 = - \frac{\LL(r-1)(r-R)(r+R+1)}{r},
$$

\noindent
and consequently the black hole's parameters will be given in our units by

\begin{align*}
& \LL = \frac1{R^2+R+1}, \\
& \MM = \frac{R^2+R}{2(R^2+R+1)}.
\end{align*}

\noindent
The (complex) tortoise coordinate which vanishes at the origin is

$$
x = \int \frac{dr}{f(r)} = \frac1{2k_H} \log(1-r) + \frac1{2k_C} \log\left(1-\frac{r}{R}\right) +\frac1{2k_F} \log\left(1+\frac{r}{R+1}\right),
$$

\noindent
where

\begin{align*}
& k_H = \frac12 f'(1) = \frac{(R-1)(R+2)}{2(R^2+R+1)}, \\
& k_C = \frac12 f'(R) = -\frac{(R-1)(2R+1)}{2R(R^2+R+1)}, \\
& k_F = \frac12 f'(-R-1) = \frac{(R+2)(2R+1)}{2(R+1)(R^2+R+1)},
\end{align*}

\noindent
are the surface gravities at the black hole horizon $r=1$, the cosmological horizon $r=R$ and the fictitious horizon $r=-R-1$. Notice that we take the surface gravity at the cosmological horizon to be negative.

As in \cite{motl-neitzke}, we notice that although $x$ has a ramification point at each horizon, ${\mathbb{R}}{\mathrm{e}} (x)$ is well defined and we can look at the Stokes line ${\mathbb{R}}{\mathrm{e}} (x)=0$. Since $x(0)=0$, this curve contains the origin and its singular points are given by

$$
\frac{dx}{dr} = 0 \quad \Leftrightarrow \quad \frac1{f(r)}=0 \quad \Leftrightarrow \quad r=0.
$$

\noindent
For $r\sim 0$ one has $f(r) \sim \frac{-2\MM}{r}$ and hence

$$
x \sim -\int \frac{rdr}{2\MM} = - \frac{r^2}{4\MM}.
$$

\noindent
Consequently the Stokes line is given by $r= \rho e^{\pm \frac{i\pi}{4}},\; \rho \in \mathbb{R}$ in a neighborhood of the origin. On the other hand, for $r \sim \infty$ one has $f(r) \sim  - \LL r^2$, and thus

$$
x \sim -\int \frac{dr}{\LL r^2} = x_0 + \frac{1}{\LL r}.
$$

\noindent
Notice that in particular $x$ has no monodromy at infinity, and hence

$$
\frac1{k_H} + \frac1{k_C} + \frac1{k_F} = 0.
$$

\noindent
Thus we can choose the three ramification lines of $x$ to cancel each other off, and $x_0$ is well defined. Using the expression for $x$ with an appropriate choice of ramification line in each logarithm, one can compute the real part of $x_0$, which is not zero. Therefore the Stokes line cannot extend all the way to infinity and the four lines starting out at the origin must thus connect among themselves. Studying the behavior of ${\mathbb{R}}{\mathrm{e}} (x)$ near the horizons, it is not hard to guess that the Stokes line is as indicated in figure 1. This guess is moreover verified by a numerical computation of the same Stokes line, indicated in figure 2.

\begin{figure}[ht] \label{StokesSdS}
    \begin{center}
      \psfrag{1}{$1$}
      \psfrag{R}{$R$}
      \psfrag{-R-1}{$-R-1$}
      \psfrag{Re}{${\mathbb{R}}{\mathrm{e}}$}
      \psfrag{Im}{${\mathbb{I}}{\mathrm{m}}$}
      \psfrag{contour}{contour}
       \psfrag{Stokes line}{Stokes line}
      \epsfxsize=.6\textwidth
      \leavevmode
      \epsfbox{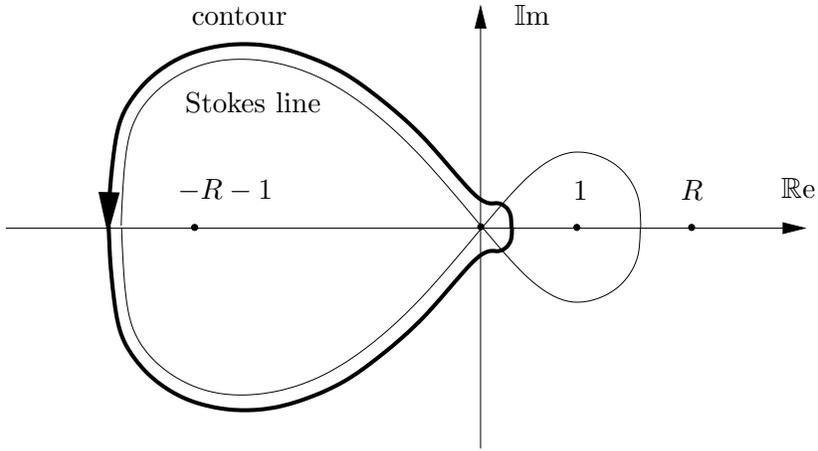}
    \end{center}
\caption{Stokes line for the Schwarzschild dS black hole, along with the chosen contour for monodromy matching.}
\end{figure}

\begin{figure}[ht] \label{SdS}
    \begin{center}
      \epsfxsize=.4\textwidth
      \leavevmode
      \epsfbox{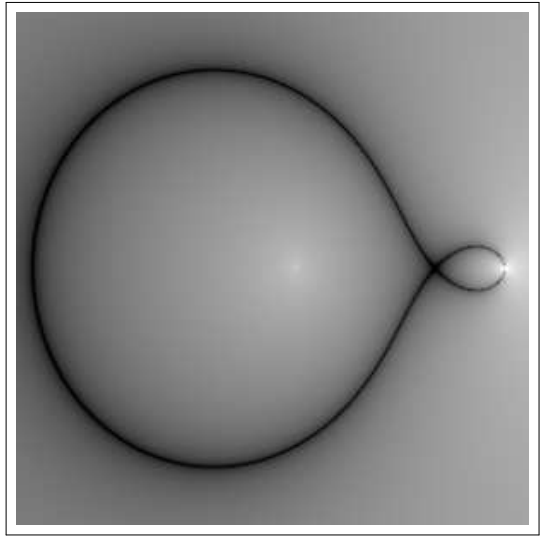}
    \end{center}
\caption{Numerical calculation of the Stokes line for the Schwarzschild dS black hole.}
\end{figure}

Since for $r \sim 0$ the presence of the cosmological constant is irrelevant, we expect the potential to behave as in the Schwarzschild black hole, and this is indeed the case:

$$
V \sim \frac{j^2-1}{4x^2},
$$

\noindent
where $j=0$ for scalar fields and scalar--type gravitational perturbations, $j=1$ for electromagnetic perturbations and $j=2$ for vector--type gravitational perturbations. Correspondingly, for $x\sim 0$ the complexified solution of the Schr\"odinger--like equation is of the form

$$
\Phi (x) \sim A_+ \sqrt{2\pi\omega x}\ J_{\frac{j}{2}} \left( \omega x \right) + A_- \sqrt{2\pi\omega x}\ J_{-\frac{j}{2}} \left( \omega x \right),
$$

\noindent
where $J_\nu$ represents a Bessel function of the first kind and $A_\pm$ are (complex) integration constants.

For the asymptotic quasinormal modes one has ${\mathbb{I}}{\mathrm{m}} (\omega) \gg {\mathbb{R}}{\mathrm{e}} (\omega)$, and hence $\omega$ is approximately purely imaginary. Consequently in a neighborhood of the origin one has $\omega x \in \mathbb{R}^+$ for $r=\rho e^{\frac{i\pi}{4}},\; \rho \in \mathbb{R}$, and $\omega x \in \mathbb{R}^-$ for $r=\rho e^{-\frac{i\pi}{4}},\; \rho \in \mathbb{R}$. From the asymptotic expansion

$$
J_\nu(z)=\sqrt{\frac{2}{\pi z}} \cos\left(z-\frac{\nu \pi}{2}-\frac{\pi}4 \right), \quad z \gg 1,
$$

\noindent
we see that

\begin{align*}
\Phi (x) & \sim 2 A_+ \cos \left( \omega x - \alpha_+ \right) + 2 A_- \cos \left( \omega x - \alpha_- \right) = \\
         & = \left( A_+ e^{-i\alpha_+} + A_- e^{-i\alpha_-}\right) e^{i \omega x} + \left( A_+ e^{i\alpha_+} + A_- e^{i\alpha_-}\right) e^{-i \omega x},
\end{align*}

\noindent
for $r=\rho e^{\frac{i\pi}{4}},\; \rho \in \mathbb{R}^-$, where

$$
\alpha_\pm = \frac{\pi}4 (1 \pm j).
$$

\noindent
For $z \sim 0$ one has the expansion

$$
J_\nu(z)=z^\nu w(z),
$$

\noindent
where $w(z)$ is an even holomorphic function. Consequently, as one rotates from $r=\rho e^{\frac{i\pi}{4}},\; \rho \in \mathbb{R}^-$ to $r=\rho e^{- \frac{i\pi}{4}},\; \rho \in \mathbb{R}^-$, one has

$$
\sqrt{2\pi e^{3\pi i} \omega x}\ J_{\pm\frac{j}{2}} \left( e^{3\pi i} \omega x \right) = e^{\frac{3\pi i}2 (1 \pm j)}\sqrt{2\pi \omega x}\ J_{\pm\frac{j}{2}} \left( \omega x \right) \sim 2 e^{6i\alpha_\pm} \cos(\omega x - \alpha_\pm)
$$

\noindent
and hence

\begin{align*}
\Phi (x) & \sim 2 A_+ e^{6i\alpha_+} \cos \left( -\omega x - \alpha_+ \right) + 2 A_- e^{6i\alpha_-} \cos \left( -\omega x - \alpha_- \right) = \\
         & = \left( A_+ e^{7i\alpha_+} + A_- e^{7i\alpha_-}\right) e^{i \omega x} + \left( A_+ e^{5i\alpha_+} + A_- e^{5i\alpha_-}\right) e^{-i \omega x},
\end{align*}

\noindent
for $r=\rho e^{\frac{-i\pi}{4}},\; \rho \in \mathbb{R}^-$. As one would expect, this completely parallels the computation for the Schwarzschild solution in \cite{motl-neitzke}. Next we compute the monodromy of the solution at infinity. For $r \sim \infty$ we have

\begin{align*}
& V_{\mathsf{s}} \sim 2 \LL^2 r^2 \sim \frac{2}{(x-x_0)^2} = \frac{3^2-1}{4(x-x_0)^2}, \\
& V_{\mathsf{em}} \sim V_{\mathsf{odd}} \sim - \LL \ell (\ell+1), \\
& V_{\mathsf{even}} \sim - \LL \left[ \ell (\ell+1) - \frac{18 \MM^{2} \LL}{a^{2}} \right].
\end{align*}

\noindent 
Consequently, either

$$
\Phi (x) \sim B_+ \sqrt{2\pi\omega (x-x_0)}\ J_{\frac{3}{2}} \left( \omega (x-x_0) \right) + B_- \sqrt{2\pi\omega (x-x_0)}\ J_{-\frac{3}{2}} \left( \omega (x-x_0) \right)
$$

\noindent
or\footnote{Here ${\widetilde{\omega}}^{2} = \omega^{2} + \LL \ell (\ell+1)$ for the electromagnetic and odd parity perturbations, and ${\widetilde{\omega}}^{2} = \omega^{2} + \LL \left[ \ell (\ell+1) - \frac{18 \MM^{2} \LL}{a^{2}} \right]$ for the even parity perturbations.}

$$
\Phi (x) \sim B_+ e^{i \widetilde{\omega} (x-x_0)} + B_- e^{-i \widetilde{\omega} (x-x_0)}
$$

\noindent
for $r \sim \infty$; in any case, $\Phi$ is holomorphic and hence the monodromy of $\Phi$ at infinity is equal to one. If $\Phi$ corresponds to a quasinormal mode, its monodromy around $r=1$ must be the same as the monodromy of $e^{i\omega x}$, that is, $e^{i\omega\frac{2\pi i}{2 k_H}} = e^{-\frac{\pi \omega}{k_H}}$. Similarly, its monodromy around $r=R$ must be the same as the monodromy of $e^{-i\omega x}$, that is, $e^{\frac{\pi \omega}{k_C}}$. Since the only other singularities of $\Phi$ are at the origin and at $r=-R-1$, it is then clear that the monodromy of $\Phi$ around the contour depicted in the figure must be

$$
\frac{1}{e^{-\frac{\pi \omega}{k_H}+\frac{\pi \omega}{k_C}}} = e^{\frac{\pi \omega}{k_H}-\frac{\pi \omega}{k_C}}.
$$

The monodromy of $e^{\pm i \omega}$ around the contour is $e^{\pm i \omega \frac{2\pi i}{2k_F}} = e^{\mp \frac{\pi \omega}{k_F}}$. As one goes around the contour the coefficient of $e^{i \omega x}$ in the asymptotic expansion of $\Phi$ gets multiplied by

$$
\frac{A_+ e^{7i\alpha_+} + A_- e^{7i\alpha_-}}{A_+ e^{-i\alpha_+} + A_- e^{-i\alpha_-}}.
$$

\noindent
For this term to have the required monodromy we must impose

$$
\frac{A_+ e^{7i\alpha_+} + A_- e^{7i\alpha_-}}{A_+ e^{-i\alpha_+} + A_- e^{-i\alpha_-}} e^{-\frac{\pi \omega}{k_F}} = e^{\frac{\pi \omega}{k_H}-\frac{\pi \omega}{k_C}} \quad \Leftrightarrow  \quad \frac{A_+ e^{7i\alpha_+} + A_- e^{7i\alpha_-}}{A_+ e^{-i\alpha_+} + A_- e^{-i\alpha_-}} = e^{-\frac{2\pi \omega}{k_C}}.
$$

\noindent
Similarly, for the term in $e^{-i \omega x}$ we get the condition

$$
\frac{A_+ e^{5i\alpha_+} + A_- e^{5i\alpha_-}}{A_+ e^{i\alpha_+} + A_- e^{i\alpha_-}} e^{\frac{\pi \omega}{k_F}} = e^{\frac{\pi \omega}{k_H}-\frac{\pi \omega}{k_C}} \quad 
\Leftrightarrow \quad \frac{A_+ e^{5i\alpha_+} + A_- e^{5i\alpha_-}}{A_+ e^{i\alpha_+} + A_- e^{i\alpha_-}} = e^{\frac{2\pi \omega}{k_H}}.
$$

\noindent
The condition for these equations to have nontrivial solutions $(A_+,A_-)$ is then

\begin{align*}
\left| 
\begin{array}{ccc}
e^{7i\alpha_+} - e^{-\frac{2\pi \omega}{k_C}} e^{-i\alpha_+} & \,\,\,\, & e^{7i\alpha_-} - e^{-\frac{2\pi \omega}{k_C}} e^{-i\alpha_-} \\ & & \\
e^{5i\alpha_+} - e^{\frac{2\pi \omega}{k_H}} e^{i\alpha_+}   &  & e^{5i\alpha_-} - e^{\frac{2\pi \omega}{k_H}} e^{i\alpha_-}
\end{array}
\right| = 0 
\quad \Leftrightarrow \quad
\left| 
\begin{array}{ccc}
\sin\left( 4\alpha_+ - \frac{i\pi\omega}{k_C}\right) & & \sin\left( 4\alpha_- - \frac{i\pi\omega}{k_C}\right) \\ & & \\
\sin\left( 2\alpha_+ + \frac{i\pi\omega}{k_H}\right) & & \sin\left( 2\alpha_- + \frac{i\pi\omega}{k_H}\right)
\end{array}
\right| = 0.
\end{align*}

\noindent
As in the Schwarzschild case, this equation is automatically satisfied for $j=0$. This is to be expected, as for $j=0$ the Bessel functions $J_{\pm\frac{j}2}$ coincide and do not form a basis for the space of solutions of the Schr\"odinger--like  equation near the origin. As in \cite{motl-neitzke}, we consider this equation for $j$ nonzero and take the limit as $j\to 0$. This amounts to writing the equation as a power series in $j$ and equating to zero the first nonvanishing coefficient, which in this case is the coefficient of the linear part. Thus, we just have to require that the derivative of the determinant above with respect to $j$ be zero for $j=0$. This amounts to

\begin{align*}
& \left|
\begin{array}{ccc}
\pi \cos\left( \pi - \frac{i\pi\omega}{k_C}\right) & & -\pi \cos\left( \pi - \frac{i\pi\omega}{k_C}\right) \\ & & \\
\sin\left( \frac{\pi}2 + \frac{i\pi\omega}{k_H}\right) & & \sin\left( \frac{\pi}2 + \frac{i\pi\omega}{k_H}\right)
\end{array}
\right|+
\left|
\begin{array}{ccc}
\sin\left( \pi - \frac{i\pi\omega}{k_C}\right) & & \sin\left( \pi - \frac{i\pi\omega}{k_C}\right) \\ & & \\
\frac{\pi}2 \cos\left( \frac{\pi}2 + \frac{i\pi\omega}{k_H}\right) & & -\frac{\pi}2 \cos\left( \frac{\pi}2 + \frac{i\pi\omega}{k_H}\right)
\end{array}
\right| = 0,
\end{align*}

\noindent
from where we obtain our final result as

\begin{equation}\label{resultsds}
\cosh\left( \frac{\pi\omega}{k_H}-\frac{\pi\omega}{k_C}\right) + 3 \cosh\left( \frac{\pi\omega}{k_H}+\frac{\pi\omega}{k_C}\right)=0.
\end{equation}

\noindent
Notice that if $\omega$ is a solution of this equation then so is $-\bar{\omega}$, as must be the case with quasinormal modes.

To recover the Schwarzschild quasinormal frequencies we first write out the equation as

$$
e^{\frac{\pi\omega}{k_H}-\frac{\pi\omega}{k_C}} + e^{-\frac{\pi\omega}{k_H}+\frac{\pi\omega}{k_C}} + 3 e^{ \frac{\pi\omega}{k_H}+\frac{\pi\omega}{k_C}} + 3 e^{ -\frac{\pi\omega}{k_H}-\frac{\pi\omega}{k_C}} = 0,
$$

\noindent
and next take the limit as $R \to \infty$, in which $k_H \to \frac12$ and $k_C \to 0^-$. If we assume that ${\mathbb{R}}{\mathrm{e}} (\omega) > 0$, we see that the two middle terms are exponentially small, and hence the equation reduces to

$$
e^{\frac{\pi \omega}{k_H}}+3e^{-\frac{\pi \omega}{k_H}}=0 \quad \Leftrightarrow \quad e^{4\pi\omega}=-3,
$$

\noindent
which is exactly the equation obtained in \cite{motl, motl-neitzke}. Therefore, the Schwarzschild black hole is not a singular limit of the Schwarzschild dS black hole as far as the quasinormal modes are concerned, unlike what happens with the Reissner--Nordstr\o m black hole solution. The reason for this is clear from the monodromy calculation: whereas the structure of the tortoise near the singularity $r=0$ in the Reissner--Nordstr\o m solution depends crucially on whether the charge is zero or not, in the Schwarzschild dS case it does not depend on $\LL$. Thus, as $R\to+\infty$, the cosmological horizon approaches the point at infinity and the contour approaches the contour used in \cite{motl-neitzke}.

For $j=1$ one has $\alpha_+=\frac{\pi}2$, $\alpha_-=0$ and hence the condition for the quasinormal frequencies is

$$
\left| 
\begin{array}{ccc}
-\sin\left(\frac{i\pi\omega}{k_C}\right) & & -\sin\left(\frac{i\pi\omega}{k_C}\right) \\ & & \\
-\sin\left(\frac{i\pi\omega}{k_H}\right) & & \sin\left(\frac{i\pi\omega}{k_H}\right)
\end{array}
\right| = 0
\quad \Leftrightarrow \quad
\sin\left(\frac{i\pi\omega}{k_C}\right)\sin\left(\frac{i\pi\omega}{k_H}\right)=0,
$$

\noindent
with the solutions

$$
\omega = ink_H \quad \mathrm{or} \quad \omega = ink_C \quad  (n\in\mathbb{N}).
$$

\noindent
Again, as $R \to \infty$ one obtains the Schwarzschild result, $\omega = \frac{ni}2$. Finally, for $j=2$ one has $2\alpha_\pm=\frac{\pi}2 \pm \pi$, $4\alpha_\pm=\pi \pm 2\pi$, and consequently the quasinormal frequencies are the same as in the $j=0$ case, for which $2\alpha_\pm=\frac{\pi}2$, $4\alpha_\pm=\pi$.

Let us now review the literature concerning asymptotic quasinormal frequencies in the Schwarzschild dS spacetime, so that we can compare our results to what has been previously accomplished on this subject. First of all, it is possible to prove, without computing explicitly the quasinormal frequencies, that $j=0$ and $j=2$ perturbations must have the same quasinormal spectra \cite{chandra}, and so this is a consistency check on our results. For Schwarzschild dS, early results on quasinormal modes were studied in \cite{bclp}, without great emphasis on the asymptotic case. The first analytical results in $d=4$ were derived in the near--extremal situation, where event and cosmological horizons are nearly coincident \cite{cardoso-lemos}, but the approximation used therein is not expected to hold in the asymptotic limit, at least on what concerns the real part of the asymptotic frequencies. Further approximations were studied in \cite{suneeta}, in a limit where the black hole mass is much smaller than the spacetime radius of curvature, but focusing explicitly on the time \textit{dependent} transient situation. An attempt at an analytic solution for the asymptotic quasinormal frequencies, using the monodromy technique of \cite{motl-neitzke}, was done in \cite{castellobranco-abdalla}. However, an erroneous identification of the relevant contours led these authors to an incorrect result (see also \cite{medved-martin}, where other arguments were given trying to explain the failure of \cite{castellobranco-abdalla} to reproduce available numerical data). Perhaps the most thorough analytical work on Schwarzschild dS asymptotic quasinormal frequencies to date is the one in \cite{choudhury-padmanabhan}. These authors find that because there are two different surface gravities, there are also two sets of solutions for ${{\mathbb{I}}{\mathrm{m}}} \left( \omega \right)$ when the horizons are \textit{widely} spaced, namely ${{\mathbb{I}}{\mathrm{m}}} \left( \omega \right)$ equally spaced with spacing equal to $k_{H}$ or ${{\mathbb{I}}{\mathrm{m}}} \left( \omega \right)$ equally spaced with spacing equal to $k_{C}$. It was further claimed in \cite{choudhury-padmanabhan} that this lack of consensus on quasinormal frequencies was due to the fact that there is no global definition of temperature in this spacetime. Our results appear to confirm this expectation: in the limit where the cosmological radius goes to infinity, and one recovers the Schwarzschild modes, we found the spacing to be equal to $k_{H}$. In the limit where the black hole radius is very small, (\ref{resultsds}) yields modes with spacing equal to $k_{C}$ (notice that this formula does not depend on the choice of units).

Besides the mentioned works, there are also numerical results available, and this is where our comparisons prove to be most conclusive. In \cite{yoshida-futamase} the asymptotic quasinormal frequencies for electromagnetic and gravitational perturbations of nearly extremal Schwarzschild dS spacetimes were studied. It turned out that, for gravitational perturbations, the real part of the asymptotic quasinormal frequencies has an oscillatory behavior as plotted against its imaginary part. We plot the same figure as in \cite{yoshida-futamase}, using our final result (\ref{resultsds}), in figure 3. This figure corresponds to the roots of (\ref{resultsds}) for a near extremal black hole with $k_{H} = 10^{-3}$, and should be compared to figure 4 in \cite{yoshida-futamase} which refers to the same value of $k_{H}$. One immediately observes agreement to large accuracy: first, the oscillation period is exactly the same. Second, the value of ${\mathrm{max}} \left( {\mathbb{R}}{\mathrm{e}} ( \omega ) \right)$ in \cite{yoshida-futamase} is of the same order as our maximum, but always larger than it, as it must be since our value refers to the asymptotic regime only. Moreover, and on what concerns $j=1$ electromagnetic perturbations, the numerical data in \cite{yoshida-futamase} is very clear and indicates that the real part of the asymptotic quasinormal frequencies should vanish, and this is precisely what we have obtained. Further numerical results have recently been obtained in \cite{konoplya-zhidenko}, this time around without any near extremality constraints. Again, for the gravitational perturbations, the real part of the asymptotic quasinormal frequencies was found to have an oscillatory behavior. Also, for electromagnetic perturbations, the real part of the asymptotic quasinormal frequencies was found to vanish. Our analytical results still agree very well with the numerics: we are able to reproduce the basic features of figure 2 in \cite{konoplya-zhidenko}, with the exception that we also find modes with a zero real part. This should not be cause for concern as it is known to be highly difficult to numerically obtain modes with a vanishing real part. On what concerns the $j=1$ perturbations of \cite{konoplya-zhidenko}, we obtain exactly what they have found.

\begin{figure}[ht] \label{SdSnumerics}
    \begin{center}
      \psfrag{Im}{${\mathbb{I}}{\mathrm{m}}$}
      \epsfxsize=.4\textwidth
      \leavevmode
      \epsfbox{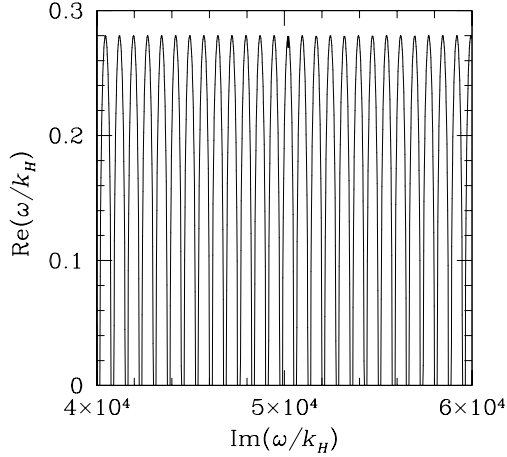}
    \end{center}
\caption{Real versus imaginary part of quasinormal frequencies for the nearly extremal Schwarzschild dS solution.}
\end{figure}


\section{Asymptotically Anti--de Sitter Spacetimes}


\cite{kodama-ishibashi-3} discusses the stability of black holes in asymptotically Anti--de Sitter (AdS) spacetimes to tensor, vector and scalar perturbations of the metric and the electromagnetic field. For black holes without charge, which is the case we shall focus on, tensor and vector perturbations are stable in any dimension. Scalar perturbations are stable in dimension four but there is no proof of stability in dimension $d \ge 5$. As we shall work in four dimensions, we are guaranteed a stable solution. Quantization of scalar field in AdS was first addressed in \cite{ais}. An important theme concerned boundary conditions: because AdS light rays can reach spatial infinity and return to the origin in finite time, as measured by the observer at the origin, one could expect for reflecting boundary conditions at the AdS walls. However, these walls are at timelike spatial infinity. As it turns out, the sensible boundary conditions to impose on quasinormal modes include the usual incoming waves at the black hole horizon and then vanishing of the fields at infinity (\textit{i.e.}, at the boundary of AdS). The Schwarzschild AdS solution in dimension $d=4$ has parameters $\MM$ and $\LL < 0$, with metric

$$
f(r) = 1 - \frac{2\MM}{r} - \LL r^{2} = 1 - \frac{2\MM}{r} + | \LL | r^{2}.
$$

\noindent
The potentials to be used in the master equation (\ref{schrodinger}) are the same as before. Also, in the following we shall focus on scalar field perturbations.

Again, to simplify the calculation we choose the radius of the black hole to be our length unit. The length scale determined by the cosmological constant will then be an adimensional quantity $R$, with $\LL = -\frac1{R^2}$. It is then easily seen that the warp factor must be of the form

$$
f(r)=1 - \frac{2\MM}{r} - \LL r^2 = \frac{(r-1)(r^2+r+1+R^2)}{R^2r}
$$

\noindent
and consequently the black hole mass will be given in our units by

$$
\MM = \frac{1+R^2}{2R^2}.
$$

\noindent
Moreover, one can compute

$$
f(r)=\frac{(r-1)(r-\gamma)(r-\bar{\gamma})}{R^2r},
$$

\noindent
where

$$
\gamma = -\frac12 + \frac{i}2 \sqrt{4R^2+3}.
$$

\noindent
For simplicity, we shall consider quasinormal modes for large black holes only, in which $R \ll 1$. This serves our purpose of illustrating how the techniques in \cite{motl-neitzke} are generalized for asymptotically AdS spacetimes, whereas the full case is carefully analyzed in \cite{natario-schiappa}. For these large black holes we have, approximately,

$$
\gamma = e^{i\frac{2\pi}3} = \sqrt[3]{1}, \quad \bar{\gamma} = \gamma^2,
$$

\noindent
and consequently

$$
\frac1{f(r)} = \frac{R^2}3 \left( \frac1{r-1} + \frac{\bar{\gamma}}{r-\gamma} + \frac{\gamma}{r-\bar{\gamma}} \right).
$$

\noindent
The (complex) tortoise coordinate which vanishes at the origin is therefore

$$
x = \int \frac{dr}{f(r)} = \frac{R^2}3 \left( \log(1-r) + \bar{\gamma} \log(1-\bar{\gamma}r) + \gamma\log(1-\gamma r) \right).
$$

We now wish to examine the Stokes line, \textit{i.e.}, the curve ${\mathbb{I}}{\mathrm{m}}(\omega x)=0$. However, this time around, the expression of the tortoise makes it clear that ${\mathbb{I}}{\mathrm{m}}(\omega x)$ is a multivalued function. To bypass this problem, we choose a particular branch and simply trace out the curve shifting the ramification lines so that it never hits them. Note that the behavior of $e^{\pm i\omega x}$ will still be oscillatory along the curve. Again the Stokes line has a unique singular point at the origin, where four lines meet, as

$$
\omega x \sim -\int \frac{\omega rdr}{2\MM} = - \frac{\omega r^2}{4\MM}
$$

\noindent
for $r \sim 0$. To understand its behavior near the singularities, we notice that following our procedure, the curve

$$
{\mathbb{I}}{\mathrm{m}} \left( \alpha \log (z) \right) = 0
$$

\noindent
is the curve

$$
\alpha \log(z) = \rho \quad \Leftrightarrow \quad z = e^{\xi \rho} e^{i \eta \rho}
$$

\noindent
(with $\alpha = \frac1{\xi + i \eta}$ and $\rho \in \mathbb{R}$ a parameter). This is a spiral that approaches the singularity $z=0$, except in the case where $\xi = 0$, \textit{i.e.}, when $\alpha$ is purely imaginary. Therefore, generically one expects the Stokes line to hit all three singularities, and hence the fourth line starting out at the origin must extend all the way to infinity. For $r \sim \infty$ we have

$$
x \sim \int \frac{R^2dr}{r^2} = x_0 - \frac{R^2}{r}.
$$

\noindent
In particular, $x$ has no monodromy at infinity, as can also be seen from its expression and the fact that

$$
1+\gamma+\gamma^2=0.
$$

\noindent
We can therefore choose the three ramification lines of $x$ to cancel each other off, making $x_0$ well defined\footnote{There are however three nonequivalent ways of doing this, leading to three possible values of $x$ at infinity: $x_0$, $\bar{x}_0$ and $-|x_0|$. The second choice leads to the quasinormal frequencies $-\bar{\omega}$, where $\omega$ are the quasinormal frequencies obtained by choosing $x_0$; the third choice leads to no solutions.}. Using the expression for $x$ with appropriate choice of ramification line in each logarithm, one then obtains

$$
x_0 = \frac{2\pi \sqrt{3}R^2}9 e^{-\frac{i\pi}3}.
$$

\noindent
Therefore we must have

$$
\omega x_0 \in \mathbb{R} \quad \Leftrightarrow \quad \omega = \zeta e^{\frac{i\pi}3} \quad (\zeta \in \mathbb{R}^+).
$$

\noindent
For such a value of $\omega$ and our choice of ramification lines, it is easily seen that $\omega x$ is real for $r=\rho e^{\frac{i\pi}3},\; \rho \in (-1,+\infty)$. On the other hand, near the origin the Stokes line is given by

$$
{\mathbb{I}}{\mathrm{m}}\left(-e^{\frac{i\pi}3} r^2 \right)=0 \quad \Leftrightarrow \quad r = \rho e^{-\frac{i\pi}6} \quad \mathrm{or} \quad r = \rho e^{\frac{i\pi}3} \quad (\rho \in \mathbb{R}).
$$

\noindent
Consequently, it is not hard to guess that the Stokes line is as depicted in figure 4. We have verified this guess with a numerical computation of the same Stokes line, and this is indicated in figure 5. It should be noted that due to the ramification lines (which can be readily identified in the figure) the spirals at the singularities are not so clearly depicted in the numerical result.

\begin{figure}[ht] \label{StokesSAdS}
    \begin{center}
      \psfrag{1}{$1$}
      \psfrag{g}{$\gamma$}
      \psfrag{g2}{$\gamma^2$}
      \psfrag{Re}{${\mathbb{R}}{\mathrm{e}}$}
      \psfrag{Im}{${\mathbb{I}}{\mathrm{m}}$}
      \psfrag{Stokes line}{Stokes line}
      \epsfxsize=.6\textwidth
      \leavevmode
      \epsfbox{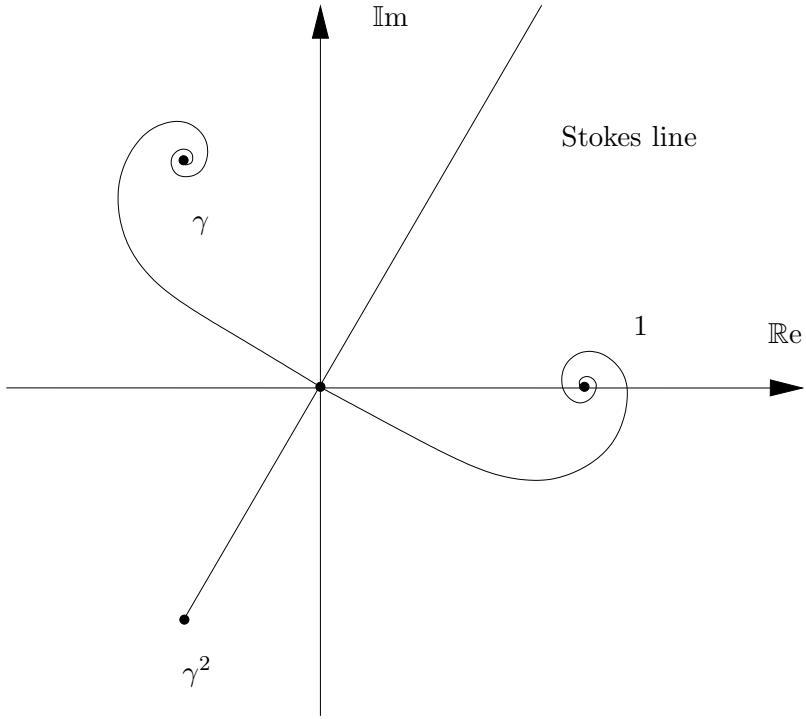}
    \end{center}
\caption{Stokes line for the Schwarzschild AdS black hole.}
\end{figure}

\begin{figure}[ht] \label{SAdS}
    \begin{center}
      \epsfxsize=.4\textwidth
      \leavevmode
      \epsfbox{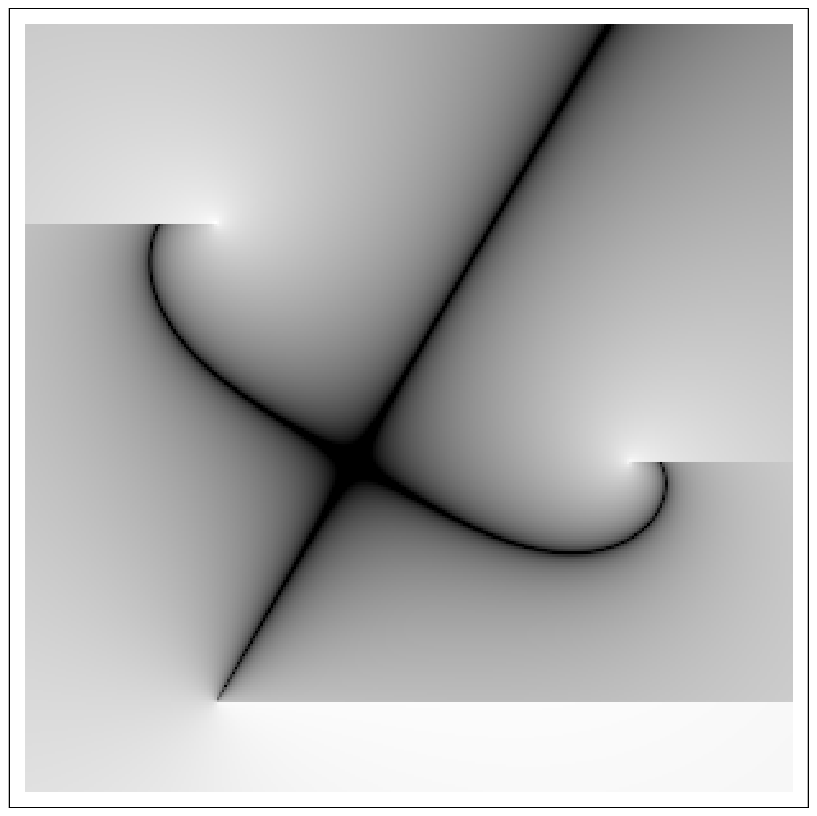}
    \end{center}
\caption{Numerical calculation of the Stokes line for the Schwarzschild AdS black hole.}
\end{figure}

Now, for $r \sim \infty$ we have 

$$
V_{\mathsf{s}} \sim 2 \LL^2 r^2 \sim \frac{2}{(x-x_0)^2} = \frac{3^2-1}{4(x-x_0)^2}.
$$

\noindent 
Consequently,

$$
\Phi (x) \sim B_+ \sqrt{2\pi\omega (x-x_0)}\ J_{\frac{3}{2}} \left( \omega (x-x_0) \right) + B_- \sqrt{2\pi\omega (x-x_0)}\ J_{-\frac{3}{2}} \left( \omega (x-x_0) \right)
$$

\noindent
for $r \sim \infty$. The boundary condition $\Phi = 0$ at $r = \infty$ requires that $B_-=0$. Hence,

$$
\Phi (x) \sim B_+ \sqrt{2\pi\omega (x-x_0)}\ J_{\frac{3}{2}} \left( \omega (x-x_0) \right)
$$

\noindent
for $r \sim \infty$. Now, and as in the Schwarzschild dS case, for $r \sim 0$ the presence of the cosmological constant is irrelevant, and the potential behaves as in the Schwarzschild black hole,

$$
V_{\mathsf{s}} \sim \frac{j^2-1}{4x^2},
$$

\noindent
where $j=0$ for scalar field perturbations. Correspondingly, for $x\sim 0$ the complexified solution of the Schr\"odinger--like equation is of the form

$$
\Phi (x) \sim A_+ \sqrt{2\pi\omega x}\ J_{\frac{j}{2}} \left( \omega x \right) + A_- \sqrt{2\pi\omega x}\ J_{-\frac{j}{2}} \left( \omega x \right),
$$

\noindent
where $A_\pm$ are (complex) integration constants. One has $\omega x \in \mathbb{R}^+$ for $r=\rho e^{\frac{i\pi}{3}},\; \rho \in \mathbb{R}$, and $\omega x \in \mathbb{R}^-$ for $r=\rho e^{-\frac{i\pi}{6}},\; \rho \in \mathbb{R}$. From the asymptotic expansion for $J_\nu(z)$, with $z \gg 1$, we see that

\begin{align*}
\Phi (x) & \sim 2 A_+ \cos \left( \omega x - \alpha_+ \right) + 2 A_- \cos \left( \omega x - \alpha_- \right) = \\
         & = \left( A_+ e^{-i\alpha_+} + A_- e^{-i\alpha_-}\right) e^{i \omega x} + \left( A_+ e^{i\alpha_+} + A_- e^{i\alpha_-}\right) e^{-i \omega x},
\end{align*}

\noindent
for $r=\rho e^{\frac{i\pi}{3}},\; \rho \in \mathbb{R}^+$, where

$$
\alpha_\pm = \frac{\pi}4 (1 \pm j).
$$

\noindent
The same expansion yields

\begin{align*}
\Phi (x) & \sim  B_+ e^{-i\beta_+} e^{i \omega (x-x_0)} + B_+ e^{i\beta_+} e^{-i \omega (x-x_0)} = \\
& = -B_+ e^{-i \omega x_0} e^{i \omega x} - B_+  e^{i \omega x_0} e^{-i \omega x}
\end{align*}

\noindent
in the same limit, since

$$
\beta_+ = \frac{\pi}4 (1 + 3) = \pi.
$$

\noindent
We conclude that $A_+, A_-$ must satisfy

$$
\left( A_+ e^{-i\alpha_+} + A_- e^{-i\alpha_-}\right) e^{i \omega x_0} = \left( A_+ e^{i\alpha_+} + A_- e^{i\alpha_-}\right) e^{-i \omega x_0}.
$$

\noindent
Again for $z \sim 0$ one has the expansion

$$
J_\nu(z)=z^\nu w(z),
$$

\noindent
where $w(z)$ is an even holomorphic function. Consequently, as one rotates from $r=\rho e^{\frac{i\pi}{3}},\; \rho \in \mathbb{R}^-$ to $r=\rho e^{- \frac{i\pi}{6}},\; \rho \in \mathbb{R}^+$, one has

$$
\sqrt{2\pi e^{-\pi i} \omega x}\ J_{\pm\frac{j}{2}} \left( e^{-\pi i} \omega x \right) = e^{\frac{-\pi i}2 (1 \pm j)}\sqrt{2\pi \omega x}\ J_{\pm\frac{j}{2}} \left( \omega x \right) \sim 2 e^{-2i\alpha_\pm} \cos(\omega x - \alpha_\pm)
$$

\noindent
and hence

\begin{align*}
\Phi (x) & \sim 2 A_+ e^{-2i\alpha_+} \cos \left( -\omega x - \alpha_+ \right) + 2 A_- e^{-2i\alpha_-} \cos \left( -\omega x - \alpha_- \right) = \\
         & = \left( A_+ e^{-i\alpha_+} + A_- e^{-i\alpha_-}\right) e^{i \omega x} + \left( A_+ e^{-3i\alpha_+} + A_- e^{-3i\alpha_-}\right) e^{-i \omega x},
\end{align*}

\noindent
for $r=\rho e^{\frac{-i\pi}{6}},\; \rho \in \mathbb{R}^+$. This form of the solution can be propagated along the corresponding branch of the Stokes line which approaches the event horizon, and where we know that $\Phi (x)$ must behave as $e^{i \omega x}$. Consequently we obtain the second condition on $A_+, A_-$, as

$$
A_+ e^{-3i\alpha_+} + A_- e^{-3i\alpha_-} = 0.
$$

\noindent
The two conditions on these coefficients can only have nontrivial solutions if and only if

\begin{align*}
\left| 
\begin{array}{ccc}
e^{-3i\alpha_+} & \,\,\,\, & e^{-3i\alpha_-} \\ & & \\
e^{-i\alpha_+} e^{i \omega x_0} - e^{i\alpha_+} e^{-i \omega x_0} &  & e^{-i\alpha_-} e^{i \omega x_0} - e^{i\alpha_-} e^{-i \omega x_0}
\end{array}
\right| = 0 
\quad \Leftrightarrow \quad
\left| 
\begin{array}{ccc}
e^{-3i\alpha_+} & \,\,\,\, & e^{-3i\alpha_-} \\ & & \\
\sin\left( \alpha_+ - \omega x_0 \right) & & \sin\left( \alpha_- - \omega x_0 \right)
\end{array}
\right| = 0.
\end{align*}

\noindent
Again, this equation is automatically satisfied for $j=0$. We must thus consider $j$ nonzero and then take the limit as $j\to 0$. This amounts to writing the equation as a power series in $j$ and equating to zero the first nonvanishing coefficient, which in this case is the coefficient of the linear part. Thus we just have to require that the derivative of the determinant above with respect to $j$ be zero for $j=0$. This is

\begin{align*}
& \left|
\begin{array}{ccc}
-\frac{3i\pi}4 e^{-\frac{3i\pi}4} & & \frac{3i\pi}4 e^{-\frac{3i\pi}4} \\ & & \\
\sin\left( \frac{\pi}4 - \omega x_0 \right) & & \sin\left( \frac{\pi}4 - \omega x_0 \right)
\end{array}
\right|+
\left|
\begin{array}{ccc}
e^{-\frac{3i\pi}4} & & e^{-\frac{3i\pi}4} \\ & & \\
\frac{\pi}4 \cos \left( \frac{\pi}4 - \omega x_0 \right) & & - \frac{\pi}4 \cos \left( \frac{\pi}4 - \omega x_0 \right)
\end{array}
\right|=0
\end{align*}

\noindent
from where we obtain our final result as

\begin{equation}
\tan \left( \frac{\pi}4-\omega x_0 \right) = \frac{i}3 \quad \Leftrightarrow \quad \omega x_0 = \frac{\pi}4 - \arctan \left( \frac{i}3 \right) + n \pi \quad (n \in \mathbb{N}).
\end{equation}

\noindent
If one makes use of the notation $\omega = [\mathrm{offset}] + n[\mathrm{gap}]$ (which is slightly different from the one in the introduction), it is simple to obtain the numerical values

$$
[\mathrm{offset}] = \frac{\frac{\pi}4 - \arctan \left( \frac{i}3 \right)}{R^2 \left( \frac{\sqrt{3} \pi}{9} - \frac{i\pi}3\right)} = \frac1{R^2}(0.572975 + 0.419193i)
$$

\noindent
and

$$
[\mathrm{gap}] = \frac{9}{4\sqrt{3}R^2} + \frac{9i}{4R^2} = \frac{1}{R^{2}} (1.29904 + 2.25i),
$$

\noindent
in complete agreement with available numerical results, as we shall see in the following (recall that we have taken the radius of the black hole horizon as our length unit).

We now need to review the literature concerning the calculation of asymptotic quasinormal frequencies in the Schwarzschild AdS spacetime, in order to compare our results to earlier work done on this subject. Quasinormal modes of Schwarzschild AdS black holes were addressed in \cite{chan-mann, horowitz-hubeny, giammatteo-jing}, having a direct interpretation in terms of the dual conformal field theory, as large static AdS black holes correspond to conformal field theory thermal states. However, for what concerns us in this paper, only the first modes were computed in \cite{chan-mann, horowitz-hubeny}. Due to the AdS/CFT correspondence, this work sparkled a series of investigations on AdS asymptotic quasinormal frequencies, which naturally concentrated in the five dimensional case (see, \textit{e.g.}, \cite{starinets, nunez-starinets, musiri-siopsis, siopsis}). For the case that concerns us in here, $d=4$, the first numerical results for the asymptotic quasinormal frequencies were published in \cite{berti-kokkotas-1}. These authors found that scalar perturbations are isospectral with both odd and even parity gravitational perturbations, and they also found the existence of modes with purely imaginary frequency. Later, an extensive study of asymptotic quasinormal frequencies for Schwarzschild AdS black holes in $d=4$ was done in \cite{ckl}, and numerically produced a number which exactly matches our analytical prediction. While the authors of \cite{ckl} found that the real part of the frequency mode increases with the overtone number, $n$, in what seems to be a characteristic particular to AdS space, they also found that, for the large black holes which we have studied in the present paper, the off set is

$$
[\mathrm{offset}] = \frac{1}{R^{2}} (0.578 + 0.420 i),
$$

\noindent
and the gap is

$$
[\mathrm{gap}] = \frac{1}{R^{2}} (1.299 + 2.250 i),
$$

\noindent
in complete and precise agreement with our analytical results.


\section{Future Directions}


Having opened the way for an extension of the monodromy technique introduced in \cite{motl-neitzke} to the case of non--asymptotically flat spacetimes, and having at hand the full list of potentials for both gravitational and electromagnetic perturbations of $d$--dimensional black holes \cite{kodama-ishibashi-1, kodama-ishibashi-3}, one can now proceed and compute asymptotic quasinormal modes for $d$--dimensional black holes. This is done in the upcoming \cite{natario-schiappa}. The relevance of this calculation towards quantum gravity is dual: on the one hand one would like to test if the ideas in \cite{hod, dreyer} are universal. A positive answer would have great consequences in the theoretical development of both loop quantum gravity and string theory. On the other hand, the computation of these asymptotic quasinormal frequencies will open way to the calculation of asymptotic greybody factors for $d$--dimensional black holes which, as we have alluded at previously, may have a deeper role to play on the road to quantum gravity than the quasinormal frequencies. Indeed, it is expected that these greybody factors may yield clues on dual string theoretic microscopic descriptions of black holes, at high energies \cite{neitzke, krasnov-solodukhin}. These are themes to which we shall return in future publications.

For the moment, let us conclude with some comments on upcoming results \cite{natario-schiappa}. Besides generalizing the present results to $d$--dimensions (and removing the large black hole constraint in the AdS case), we also show how to include charge in this non--asymptotically flat situation. In what respects the charged solutions, we have also generalized the computation of asymptotic quasinormal frequencies in the four dimensional Reissner--Nordstr\o m solution, of \cite{motl-neitzke}, to $d$--dimensions, with the following result

$$
e^{\beta_{H}^{+} \omega} + \Big( 1 + 2 \cos \left( \pi j \right) \Big) + \Big( 2 + 2 \cos \left( \pi j \right) \Big) e^{- \beta_{H}^{-} \omega} = 0,
$$

\noindent
where the $\beta_{H}^{\pm}$ are the inverse temperatures at the outer and the inner horizons, and 

$$
j = \frac{d-3}{2d-5}.
$$

\noindent
This formula is valid in any dimension and for any type of gravitational or electromagnetic perturbation. In the case of the extremal Reissner--Nordstr\o m solution, where mass equals charge, we have found that the asymptotic frequencies are solutions of the expression

$$
\omega = \frac{d-3}{d-2}\ \left( \frac{d-3}{4 \pi R_{H}} \right)\ \log \left( \frac{\sin \left( \frac{5 \pi j}{2} \right)}{\sin \left( \frac{\pi j}{2} \right)} \right),
$$

\noindent
where $R_{H}$ is the radius of the extremal horizon and $j$ is as before. We have also found solutions for normal modes in $d$--dimensional pure AdS space, and solutions for quasinormal modes in pure dS space, for particular values of the dimension. A complete list of all these results will appear in \cite{natario-schiappa}.

\section*{Acknowledgements}
We would like to thank Hideo Kodama for useful correspondence, Shijun Yoshida for useful comments and specially Carlos Herdeiro for very useful discussions and comments. VC is supported in part by funds provided by the Funda\c c\~ao para a Ci\^encia e a Tecnologia, under the grant SFRH/BPD/14483/2003. JN is partially supported by FCT/POCTI/FEDER. RS is supported in part by funds provided by the Funda\c c\~ao para a Ci\^encia e a Tecnologia, under the grants SFRH/BPD/7190/2001 and FEDER--POCTI/FNU/38004/2001.

\vfill

\eject

\bibliographystyle{plain}

\end{document}